\documentclass[fleqn,usenatbib]{mnras}

\usepackage{newtxtext,newtxmath}

\usepackage[english]{babel}
\usepackage{bm}
\usepackage{graphicx}
\usepackage{physics}
\usepackage{tensor}
\usepackage[capitalise]{cleveref}
\usepackage{acro}
\usepackage[normalem]{ulem}

\DeclareAcronym{GW}{short = GW, long  = gravitational wave}
\DeclareAcronym{LVK}{short = LVK, long  = LIGO-Virgo-Kagra}
\DeclareAcronym{EM}{short = EM, long  = electromagnetic}
\DeclareAcronym{BH}{short = BH, long  = black hole}
\DeclareAcronym{CBC}{short = CBC, long  = compact binary coalescence}
\DeclareAcronym{GSHE}{short = GSHE, long  = gravitational spin Hall effect}
\DeclareAcronym{ODE}{short = ODE, long  = ordinary differential equation}
\DeclareAcronym{WKB}{short = WKB, long  = Wentzel–Kramers–Brillouin}
\DeclareAcronym{GR}{short = GR, long  = general theory of relativity}
\DeclareAcronym{BL}{short = BL, long  = Boyer-Lindquist}
\DeclareAcronym{AGN}{short = AGN, long  = active galactic nuclei}
\DeclareAcronym{SNR}{short = SNR, long  = signal-to-noise ratio}
\DeclareAcronym{GO}{short = GO, long  = geometrical optics}

\usepackage[T1]{fontenc}

\DeclareRobustCommand{\VAN}[3]{#2}
\let\VANthebibliography\thebibliography
\def\thebibliography{\DeclareRobustCommand{\VAN}[3]{##3}\VANthebibliography}

\usepackage{graphicx}	
\usepackage{amsmath}	

\title[Probing general relativistic spin-orbit coupling with gravitational waves]{Probing general relativistic spin-orbit coupling with gravitational waves from hierarchical triple systems}

\author[M. A. Oancea, R. Stiskalek and M. Zumalac\'{a}rregui]{
Marius A. Oancea,$^{1}$\thanks{E-mail: marius.oancea@univie.ac.at}
Richard Stiskalek,$^{2,3}$
and Miguel Zumalac\'{a}rregui$^{4}$
\\
$^{1}$University of Vienna, Faculty of Physics, Boltzmanngasse 5, 1090 Vienna, Austria\\
$^{2}$Astrophysics, University of Oxford, Denys Wilkinson Building, Keble Road, Oxford, OX1 3RH, UK\\
$^{3}$Universit\"{a}ts-Sternwarte, Ludwig-Maximilians-Universit\"{a}t M\"{u}nchen, Scheinerstr. 1, 81679 M\"{u}nchen, Germany\\
$^{4}$Max Planck Institute for Gravitational Physics (Albert Einstein Institute), Am M\"uhlenberg 1, D-14476 Potsdam, Germany
}

\pubyear{\the\year{}}

\begin{document}
\label{firstpage}
\pagerange{\pageref{firstpage}--\pageref{lastpage}}
\maketitle

\begin{abstract}
Wave packets propagating in inhomogeneous media experience a coupling between internal and external degrees of freedom and, as a consequence, follow spin-dependent trajectories. These phenomena, well known in optics and condensed matter physics, are referred to as spin Hall effects. Similarly, the gravitational spin Hall effect is expected to affect the propagation of gravitational waves on curved spacetimes. In this general-relativistic setup, the curvature of spacetime acts as impurities in a semiconductor or inhomogeneities in an optical medium, leading to a frequency- and polarization-dependent propagation of wave packets. In this letter, we study this effect for strong-field lensed gravitational waves generated in hierarchical triple black hole systems in which a stellar-mass binary merges near a more massive black hole. We calculate how the gravitational spin Hall effect modifies the gravitational waveforms and show its potential for experimental observation. If detected, these effects will have profound implications for astrophysics and tests of general relativity.
\end{abstract}




\section{Introduction}

In optics and condensed matter physics, the dynamics of wave packets carrying intrinsic angular momentum can generally depend on spin-orbit interactions~\citep{SOI_review,SHE_review,Rashba_SOI2015}. This mechanism describes the mutual coupling between the external (average position and momentum) and internal (spin or polarization) degrees of freedom of the wave packet and is generally responsible for the spin Hall effects~\citep{SHE_review1,SHE_review,SOI_review,SHEL_review}. These effects have been observed in several experiments~\citep{originalSHE3,originalSHE4,Hosten2008,Bliokh2008}, and have led to a broad range of applications in spintronics, photonics, metrology, and optical communications \citep{Jungwirth2012,SHEL_review,Shuoqing2022}.

Similarly, spin-orbit interactions are also predicted to affect the dynamics of wave packets in gravitational fields through the \ac{GSHE}, be it for electromagnetic~\citep{GSHE2020,Harte_2022,Frolov2020,SHE_QM1} or linearized gravitational~\citep{GSHE_GW,SHE_GW} waves propagating on curved spacetimes (see also~\cite{GSHE_reviewCQG,GSHE_review,GSHE_Dirac,Li:2022izh}). This implies a certain universality of spin Hall effects across different physical systems. The analogy that we can make between the general relativistic setup and other areas of physics is that \acp{BH} in spacetime play a role similar to impurities in a semiconductor or inhomogeneities of an optical medium. Thus, under the influence of gravity, wave packets carrying intrinsic angular momentum (as is the case with electromagnetic and gravitational waves) follow frequency- and polarization-dependent trajectories, reducing to geodesic motion only in the limit of infinite frequency, i.e. \ac{GO}. Given this frequency dependence, we expect \acp{GW} to represent the most favourable avenue for observing the \ac{GSHE}.

\Acp{GW} offer a precision probe of astrophysical phenomena~\citep{TheLIGOScientific:2014jea, Abbott:2016blz},
carrying information on strong field dynamical gravity. Due to their low frequency, the \ac{GSHE} is much less suppressed for \acp{GW} than for electromagnetic signals. A fraction of \ac{GW} sources may merge in a high-curvature region. 
Active galactic nuclei (AGNs) or globular cluster binary formation channels~\citep{OLeary:2008myb,Martinez:2020lzt,Sedda:2023big,Stone:2016wzz,Secunda:2018kar,Samsing2022,Gerosa2021} provide environments where a \ac{GW} source near another \ac{BH} can produce a detectable \ac{GSHE} signal.
As the number of recorded \ac{GW} events grows, so will the prospect of such a detection.

In this letter, we present compelling theoretical and numerical evidence for astrophysical configurations in which the \ac{GSHE} is measurable on \ac{GW} signals at the current detector sensitivity in optimal situations. We will discuss the \ac{GSHE}, its imprint on waveforms, and the prospects for detection. Our results are mainly based on a numerical ray-tracing code for the \ac{GSHE}, that we make publicly available at \citep{GSHE_code}. Further details about the numerical implementation, as well as other technical details can be found in~\citep{GSHE_lensing}.

\section{Gravitational spin Hall effect}

We investigate the lensing of \acp{GW} in hierarchical triple \ac{BH} systems, where two stellar-mass \acp{BH} merge and emit \acp{GW} in the proximity of the third, much larger \ac{BH}, which acts as a lens. We assume that the merging \acp{BH} are much smaller than the lens, so that we can use the following idealized model: the lens is represented by a fixed background Kerr \ac{BH}, and the merging \acp{BH} are treated as a static point source of \acp{GW}. The emitted \acp{GW} are treated as small metric perturbations of the background Kerr \ac{BH}, and are described by the linearized Einstein field equations.

In the \ac{GO} approximation, the propagation of \acp{GW} is described by the null geodesics of the background spacetime \citep[Sec.~35.13]{MTW}. However, this does not take into account the general relativistic spin-orbit coupling between the internal and external degrees of freedom of a wave packet. This appears as higher-order corrections to the \ac{GO} approximation~\citep{GSHE2020, GSHE_GW, Harte_2022}, resulting in frequency- and polarization-dependent wave packet propagation (\ac{GSHE}). The equations of motion that describe the \ac{GSHE} are \citep{GSHE_GW, Harte_2022}
\begin{subequations}\label{eq:GSHE_ODEs}
\begin{align}
    \dot{x}^\mu &= p^\mu + \frac{1}{p \cdot t} S^{\mu \beta} p^\nu \nabla_\nu T_\beta,\\
    \dot{x}^\nu \nabla_\nu p_\mu  &= -  \frac{1}{2} R_{\mu \nu \alpha \beta}  p^\nu S^{\alpha \beta}.
\end{align}
\end{subequations}
Here, external degrees of freedom are represented by $x^\mu(\tau)$, the worldline of the energy centroid of the wave packet, and the average wave packet momentum $p_\mu(\tau)$. The internal degree of freedom is represented by the spin tensor $S^{\alpha \beta}$, which encodes the angular momentum carried by the wave packet. The timelike vector field $T^\alpha$ is needed to fix the definition of the energy centroid of the wave packet \citep{Harte_2022} and can be related to the $4$-velocities of the source and observer \citep{GSHE_lensing}, and $R_{\mu \nu \alpha \beta}$ is the Riemann tensor. Here, we consider circularly polarized wave packets, for which the spin tensor is uniquely fixed as
\begin{equation} \label{eq:spin_tensor}
    S^{\alpha \beta} = \frac{\epsilon s}{p \cdot T} \varepsilon^{\alpha \beta \gamma \lambda} p_\gamma T_\lambda,
\end{equation}
where $s = \pm 2$, depending on the state of circular polarization, $\varepsilon$ is the Levi-Civita tensor and $\epsilon$ is a small dimensionless parameter used to keep track of the order of different terms in the high-frequency expansion leading to the above equations \citep{GSHE_GW, Harte_2022}; see also \cite[Sec.~1.5]{Maggiore_GWbook}. The wave frequency $f$ measured by an observer with $4$-velocity $T^\alpha$ is defined as $p \cdot T = - \epsilon f$. For the hierarchical triple systems considered here, we define $\epsilon$ as the ratio of the wavelength $\lambda$ of the \ac{GW} (in the rest frame of the source) and half the Schwarzschild radius $R_s$ of the background \ac{BH}:
\begin{equation}\label{eq:epsilon_SI}
	\epsilon
	= \frac{\lambda}{R_s / 2} =
	\frac{c^2 \lambda}{G M}.
\end{equation}
Since \cref{eq:GSHE_ODEs} is only valid for $\epsilon \ll 1$, we will always work in a regime where $\epsilon \leq 0.1$. In particular, the \ac{GSHE} vanishes if $\epsilon \rightarrow 0$ and~\cref{eq:GSHE_ODEs} reduce to the geodesic equations.

Equations \eqref{eq:GSHE_ODEs} and \eqref{eq:spin_tensor} are a particular case of the Mathisson-Papapetrou equations, where $\dot{x}^\mu$ and $p_\mu$ are null, and the worldline is fixed by the Corinaldesi-Papapetrou spin supplementary condition $S_{\alpha \beta} T^\beta = 0$ \citep{Harte_2022}. Furthermore, the spin-dependent correction terms in the equations can also be related to the components of a Berry curvature $2$-form on phase space \citep{GSHE2020,GSHE_GW,Harte_2022}, as is also the case for spin Hall effects in condensed matter physics \citep{Berry_CM1,Berry_CM2,SHE_review} and optics \citep{SHE_original,SOI_review}.

We model the hierarchical triple \ac{BH} system as a Kerr background \ac{BH} of mass $M$ and spin parameter $a$, together with a static point source of \acp{GW} placed close to the \ac{BH} and a distant static observer. We use~\cref{eq:GSHE_ODEs} to study the propagation of \acp{GW} between the source and the observer. The \ac{GSHE} will be seen by the observer as a time delay between the frequency and polarization components of the waveform.

\begin{figure}
    \includegraphics[width=\columnwidth]{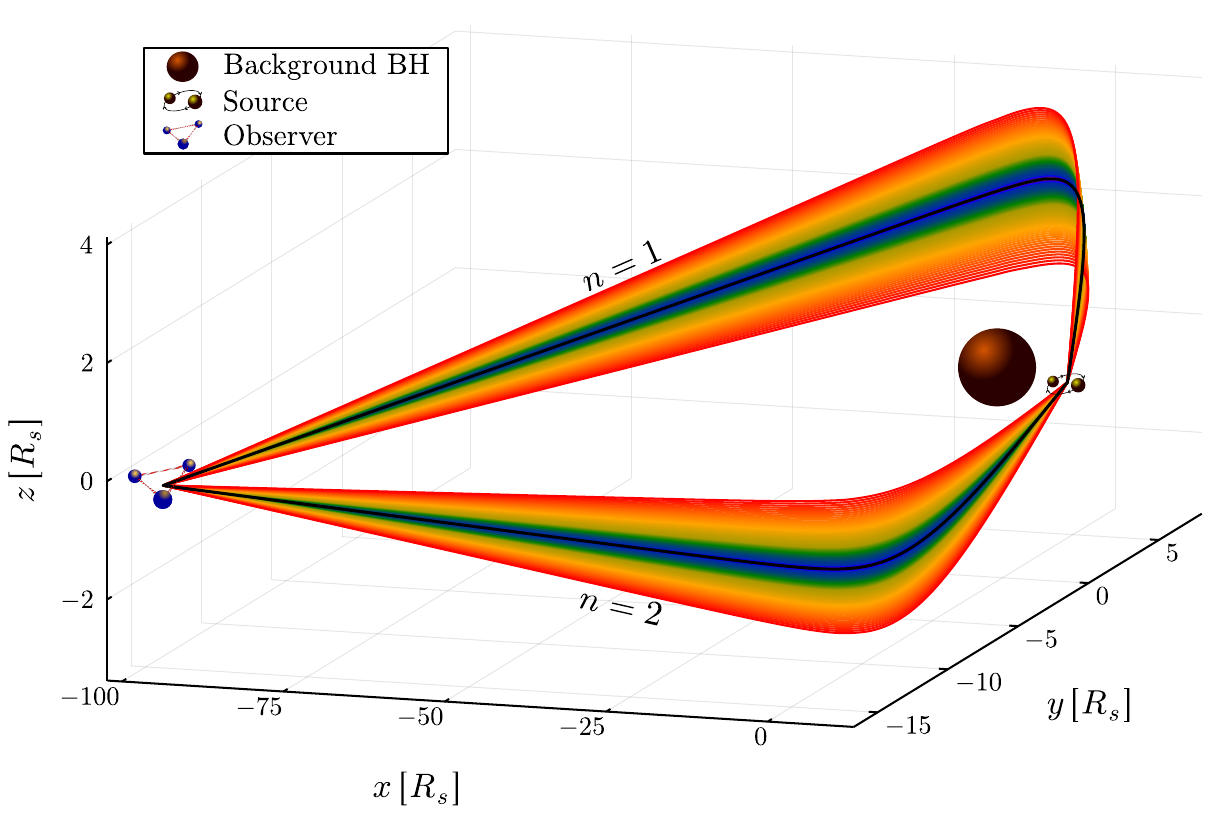}
    \caption{Two bundles of \ac{GSHE} trajectories connecting a source at $(5\,R_{\rm s}, 0.5 \pi, 0)$ to an observer at $(50\,R_{\rm s}, 0.4 \pi, \pi)$. The background \ac{BH} is Kerr with $a = 0.99 M$. Each bundle consists of a null geodesic (black trajectory in the middle of the bundle, corresponding to $\epsilon s = 0$) and $100$ \ac{GSHE} rays with $s = \pm 2$ and $\epsilon \in [10^{-3}, 10^{-1.5}]$. The trajectories are coloured according to the value of $\epsilon$, with red corresponding to longer wavelengths and violet corresponding to shorter wavelengths. Each bundle consists of two copies of a rainbow since for each finite wavelength there are two \ac{GSHE} rays of opposite circular polarization ($s = \pm 2$).}
    \label{fig:example_trajectory}
\end{figure}

In~\cref{fig:example_trajectory}, we show an example of two $\epsilon s$-parametrized bundles of trajectories that connect a source and an observer. Each bundle is centred along a null geodesic (corresponding to $\epsilon s = 0$), and we label different bundles with positive integers $n$, with $n=1$ corresponding to the shortest path. Typically, there exist two distinct bundles of trajectories that directly connect a source and an observer, and several other bundles of trajectories that loop around the \ac{BH}. We shall mainly focus on the directly connecting bundles and ignore the ones that loop around the \ac{BH}, as the latter correspond to highly demagnified signals. The connecting trajectories are determined numerically, as outlined in~\cite[Sec.~II.C]{GSHE_lensing}. This also yields the time of arrival of an $\epsilon s$-parametrized ray intersecting with the observer's worldline.

\begin{figure*}
    \centering
    \includegraphics[width=0.85\textwidth]{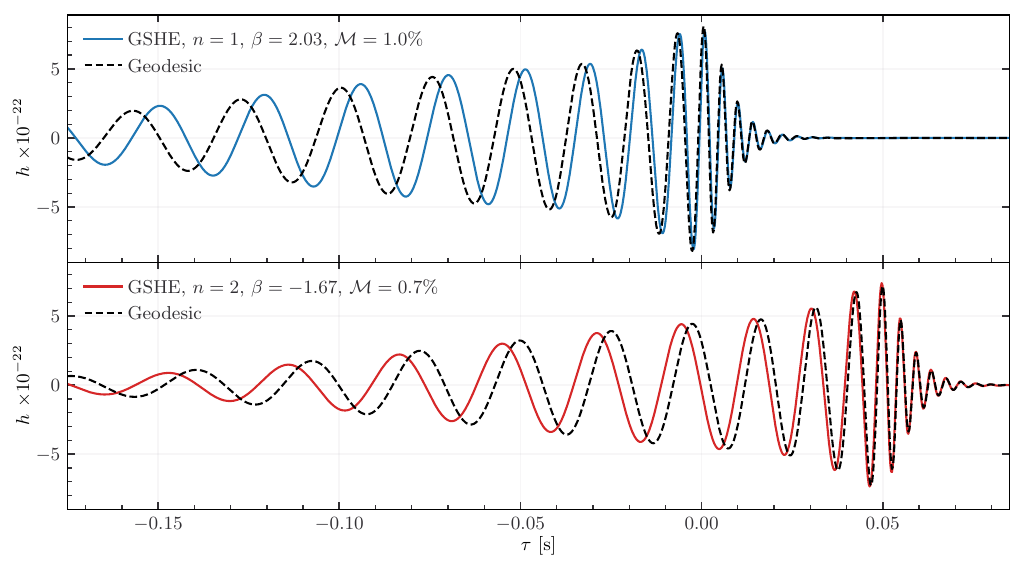}
    \caption{
    The waveform of a $50$ and $35~M_\odot$ merger propagated along the two $n$-indexed bundles shown in~\cref{fig:example_trajectory} (\emph{top} and \emph{bottom} rows). The geodesic delay between bundles is $\tau_{\rm GO}^{(2)}-\tau_{\rm GO}^{(1)}=50~\mathrm{ms}$ for this configuration. We assume $\lambda_{\max} / R_{\rm s} = 0.1$, where $\lambda_{\max}$ is the largest wavelength at $40~\mathrm{Hz}$, and report the mismatch $\mathcal{M}$. The \ac{GSHE} is a frequency-dependent phase shift in the inspiral part of the the signal.} 
    \label{fig:circ_waveform}
\end{figure*}

\section{Time delays}

The \ac{GSHE} ray propagation induces a frequency- and polarization-dependent time of arrival. The observer proper time of arrival of rays in the $n$\textsuperscript{th} bundle is denoted by $\tau_{\rm GSHE}^{\left(n\right)}(f, s)$, and we write the time delays as
\small
\begin{subequations}
\begin{align}
    \Delta \tau^{\left(n\right)} (\epsilon, s)
    &=
    \tau^{\left(n\right)}_{\rm GSHE}(\epsilon, s) - \tau^{\left(n\right)}_{\rm GO},\label{eq:tau_dispersive}\\
    \Delta \tau^{\left(n\right)}_{\rm R-L}(\epsilon)
    &=
    \tau^{\left(n\right)}_{\rm GSHE}(\epsilon, s=+2) - \tau^{\left(n\right)}_{\rm GSHE}(\epsilon, s=-2),\label{eq:tau_birrefringent}
\end{align}
\end{subequations}
\normalsize
where $\tau_{\rm GO}^{\left(n\right)}$ is the geodesic proper time of arrival. The first equation is the dispersive \ac{GSHE}-to-geodesic delay, and the second is the birefringent delay between the right- and left-polarized rays. We find that both the \ac{GSHE}-to-geodesic and right-to-left delays can be well approximated as a power law in frequency with proportionality factor $\beta$ and exponent $\alpha$ as
\begin{equation}
    \Delta \tau
    \approx
    \beta \sqrt{- g_{00}|_{\bm{x}_{\rm obs}}} \left(\frac{2c}{R_{\rm s}}\frac{1}{f}\right)^{\alpha - 1}\frac{1}{f},
\end{equation}
where $R_{\rm s} = 2 G M / c^2$ is the Schwarzschild radius of the background \ac{BH}. For the \ac{GSHE}-to-geodesic delay we denote the power law parameters by $\alpha, \beta$ and in the case of the birefringent delay by $\alpha_{\rm R-L}, \beta_{\rm R-L}$.

We find $\alpha \approx 2$ and $\alpha_{\rm R-L} \approx 3$, independently of the configuration. On the other hand, the proportionality factors $\beta$ and $\beta_{\rm R-L}$ are determined by the mutual orientation of the source and the observer with respect to the background \ac{BH} and its spin. The origin of the power law scaling and the dependence of the \ac{GSHE} on the configuration are discussed in \cite[Sec.~III.A]{GSHE_lensing}. Note that $\beta_{\rm R-L}$ is typically subdominant, but only zero in the Schwarzschild metric.

\section{Gravitational waveforms}

The \ac{GSHE}-induced time delay measured by the observer is frequency-dependent and weakly polarization-dependent. A gravitational waveform in a terrestrial detector typically spans a frequency range $2-1000~\mathrm{Hz}$ \citep{PhysRevD.102.062003,Hild_2011,Abbott_2017,Reitze:2019iox} and therefore its frequency components are delayed -- either positively or negatively, depending on the sign of $\beta$ -- with respect to the original waveform emitted by the source. The frequency components of the unlensed waveform $\tilde{h}_{0}(f, s)$ are phase shifted so that in the circular polarization basis the \ac{GSHE}-corrected waveform is
\begin{equation}\label{eq:GSHEwaveform_frequency}
\begin{split}
    \tilde{h}_{\rm GSHE}(f, s) =
    \sum_n e^{-2\pi i f \tau^{(n)}_{\rm GSHE}(f, s)}  \sqrt{\left|\mu^{(n)}(f, s)\right|} \tilde{h}_{0} (f, s).
\end{split}
\end{equation}
The sum runs over the different images, that is, the bundles that connect the source and the observer. The magnification factor $\mu^{(n)}$ has a negligible dependence on $f$ and $s$, so we will use its \ac{GO} limit.

In~\cref{fig:circ_waveform}, we show an example of the \ac{GSHE}-induced frequency-dependent delay on an \texttt{IMRPhenomXP}~\citep{IMRPhenomXP} waveform of a $50$ and $35~M_\odot$ binary \ac{BH} merger. The merger frequency is $\sim 225~\mathrm{Hz}$, we set the lower frequency limit to $40~\mathrm{Hz}$ and the background \ac{BH} mass $M = 5\times 10^4 M_\odot$. In this case, the maximum value of $\epsilon$ is $0.1$, and the \ac{GSHE}-to-geodesic delay is
\begin{equation}
    \Delta \tau \approx 3~\mathrm{ms}~\beta \left(\frac{5\times 10^4 M_\odot}{M}\right)\left(\frac{40~\mathrm{Hz}}{f}\right)^2.
\end{equation}
Due to the inverse quadratic scaling with frequency, the delay of the merger components is $\sim 30$ times less than that of the early inspiral at $40~\mathrm{Hz}$. The \ac{GSHE} introduces a frequency-dependent phase shift in the inspiral part of the waveform, which is analogous to a non-zero graviton mass if $\beta > 0$~\cite[Sec.~IV.E]{GSHE_lensing}.

As a measure of distinguishability of the \ac{GSHE} imprint on the waveform, we calculate the mismatch $\mathcal{M}$ between $\tilde{h}_{\rm GSHE}$ and the corresponding \ac{GO} signal (optimized over the coalescence phase and time), assuming a flat detector sensitivity~\cite[Sec.~II.E]{GSHE_lensing}. In~\cref{fig:circ_waveform}, for the configuration given in~\cref{fig:example_trajectory}, we find that for the two bundles $\beta \approx 2~\mathrm{and}~-1.7$ and thus $\mathcal{M} \approx 1~\%$. The \ac{GSHE} is clearly distinguishable even for a moderate \ac{SNR}~\citep{Lindblom2008}. We further assess detectability using the equivalence of the \ac{GSHE} (in the limit $\Delta\tau_{\rm L-R}\sim 0$) to tests of the modified dispersion relation for \acp{GW}, as both predict a phase shift $\propto 1/f$ on the waveform. Posterior samples of the analysed \ac{LVK} events \citep{LIGOtgrGWTC1,LIGOScientific:2020tif,LIGOScientific:2021sio} translate into $\sim\mathcal{O}\left(10^{-2}\right)$ 90\% c.l. limits on $|\beta|$, assuming $M=5 \times 10^4M_\odot$~\cite[Sec.~II.D]{GSHE_lensing}, in good agreement with the mismatch criterion. The comparison shows no strong degeneracies between \ac{GSHE} and quasi-circular binary parameters. Additional effects can be included in $\tilde h_0$ (Eq.~\ref{eq:GSHEwaveform_frequency}): Based on the different phase evolutions, we expect eccentricity~\citep{Tiwari:2019jtz} and environmental effects~\citep{Toubiana:2020drf,Sberna:2022qbn} to be distinguishable from the \ac{GSHE}. Microlensing by stellar fields can also be distinguished, as it causes stochastic variations on the phase and amplitude that oscillate in frequency~\citep{Diego:2019lcd,Mishra:2021xzz}, in contrast to the monotonic frequency-dependent phase shift associated with the GSHE.

\section{Detectability}

\begin{figure}
    \centering
    \includegraphics[width=0.97\columnwidth]{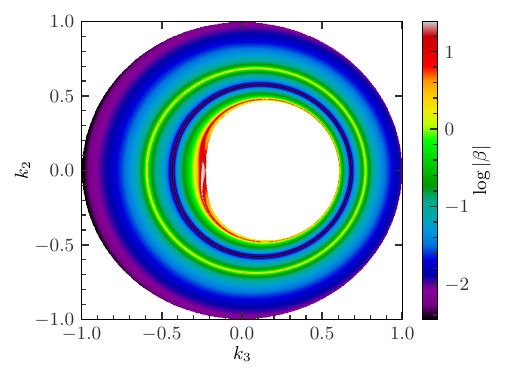}
    \caption{The \ac{GSHE}-to-geodesic delay magnitude $\beta$ as a function of the initial emission directions from the source computed with $\epsilon_{max} = 0.01$. These are parametrized by Cartesian coordinates $k_2$ and $k_3$ on the celestial sphere of the source, and the central white region represents the background \ac{BH} shadow. The source is placed at $(5\,R_{\rm s}, \pi/2, 0)$ and the ``observer'' is defined as the point where the $\epsilon_{max}$ trajectory intersects the sphere of radius $50\,R_{\rm s}$. Each pixel represents an $\epsilon$ bundle of trajectories.}
    \label{fig:shadow_beta}
\end{figure}

Throughout this work, we have assumed a fiducial background \ac{BH} mass of $5\times10^4~M_\odot$. In this regime, the wavelength of \acp{GW} detectable by terrestrial observatories is sufficiently large to deviate from the \ac{GO} propagation without requiring a wave optics treatment~\citep{Tambalo:2022plm, Leung:2023lmq} -- the regime in which our \ac{GSHE} calculation applies. We identify two favourable configurations that yield $|\beta| \gtrsim 1$: aligned source and observer (\cref{fig:example_trajectory}) and non-aligned source-observer, where a strongly deflected trajectory grazes the shadow of the background \ac{BH}. Both configurations are apparent in~\cref{fig:shadow_beta}, where the dispersive \ac{GSHE} amplitude is shown as a function of the emission direction for a source at $(5~R_{\rm s}, \pi/2, 0)$. The outer ring of $|\beta| \gtrsim 1$ corresponds to magnified bundles of trajectories toward observers closely aligned with the source-\ac{BH} system. The inner region around the \ac{BH} shadow boundary consists of bundles that are strongly deflected or even loop around the \ac{BH}. These trajectories reach non-aligned observers but are highly demagnified.

We calculate that in the scenario of~\cref{fig:shadow_beta}, approximately $5\%$ of the initial directions on the celestial half-sphere of the source facing the \ac{BH} yield $|\beta| \gtrsim 0.5$, further scaling as the inverse square of the radial source distance from the background \ac{BH}. Translating probabilities to the observer frame introduces a Jacobian element $|\mu|^{-1}$. This reflects how magnified images require a precise source-\ac{BH}-observer alignment, while demagnified images are generic: there is at least one strongly deflected trajectory that grazes the light ring and reaches any observer. Trajectories with $|\mu|\ll 1$ and $|\beta| \gtrsim 1$ are the main contributors to probability, even when demagnification is taken into account.

\begin{figure}
 \includegraphics[width=\columnwidth]{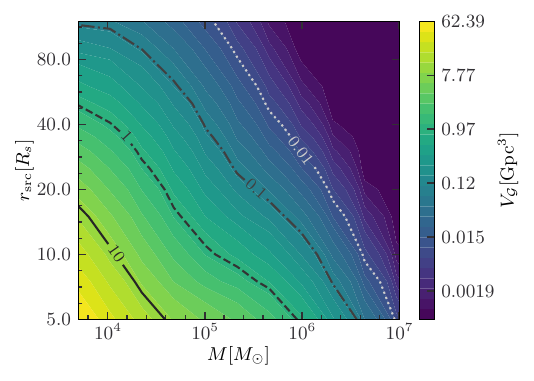}
 \caption{Effective volume for a $30+30~M_\odot$ non-spinning binary observed by Cosmic Explorer, as a function of the distance to the background \ac{BH} and its mass.}
 \label{fig:observation}
\end{figure}

To estimate the detection probabilities, we define the effective \ac{GSHE} observable volume
\begin{equation}
V_{\mathcal{G}} = \int \dd z \frac{dV_z}{dz}(z) \int \dd|\mu| P_{\rm det}  \frac{\dd \Upsilon_{\rm obs}}{\dd |\mu|},
\end{equation}
via an integral over source redshift and magnification of the product of comoving differential volume $\dd V_z / \dd z$, detected fraction $P_{\rm det}$~\citep{Chen:2017wpg} and probability of observable \ac{GSHE} in the observer sphere $\dd \Upsilon_{\rm obs} / \dd |\mu|$, both depending on the sources' properties and \ac{SNR}. \cref{fig:observation} shows $V_{\mathcal{G}}$ for quasi-circular, non-spinning $30+30M_\odot$ binary coalescences observed by Cosmic Explorer~\citep{Reitze:2019iox} as a function of the mass of the background \ac{BH} and its distance to the source (see \cite[Sec.~IV.E]{GSHE_lensing} for details). The detection rate is $\dot N_{\rm obs}\approx \mathcal{R} V_{\mathcal{G}}$, where the merger rate $\mathcal{R}(M,r_{\rm src})$ is assumed to be constant.

Configurations where the \ac{GSHE} is detectable may be realized in dense dynamical environments, such as globular and nuclear star clusters~\citep{OLeary:2008myb,Martinez:2020lzt,Sedda:2023big}. These regions contain stellar-mass \acp{BH} and may also host intermediate-mass \acp{BH}: stellar-mass binaries may then merge close to a more massive object, either by chance (among the $\sim 10^5$/yr mergers expected by future detectors) or because of its effects on the binary (e.g. if tidal interactions drive the merger as the binary approaches the intermediate-mass \ac{BH}). In a favourable case, $r_{\rm src} \sim 5~R_{s}$ and $M \sim 5 \times 10^4 M_\odot$, next-generation detectors reach an effective volume $\sim 30~{\rm Gpc}^3$. Another potential scenario consists of stellar-mass binary \acp{BH} in AGNs~\citep{Stone:2016wzz,Secunda:2018kar,Samsing2022}. There, \acp{BH} are expected to migrate inward due to interactions with the gas~\citep{Bellovary:2015ifg,Grishin:2023riv} and become trapped close to the innermost stable circular orbit of the background \ac{BH}~\citep{Peng:2021vzr}. In this case, the high mass of the background \ac{BH} suppresses the amplitude in the frequency band of ground detectors. Nevertheless, Fig.~\ref{fig:observation} shows how Cosmic Explorer alone could detect GSHE in a binary near a $10^7M_\odot$ \ac{BH} at a characteristic distance $V_{\mathcal{G}}^{1/3}\sim 200$ Mpc.
Our estimate is conservative in this limit because our simulations do not resolve well the high $\beta$ regime, which dominates the probabilities for large $M$.

The \ac{GSHE} is a promising probe of the \ac{BH} merger environment due to the frequency-dependent time delay. In addition, we expect to receive multiple, shortly spaced, images of the same merger along various bundles connecting the source and observer. The delay between the images and their relative magnification can be used to retrieve information about the \ac{BH} mass and the orientation of the source-\ac{BH}-observer system. 
The \ac{GSHE} provides additional information, including a direct constraint on the spin of the background \ac{BH} if the birefringence effect $\Delta \tau_{\rm R-L}$ is observed.
Moreover, neglecting the \ac{GSHE} or the interference between multiple images can prevent detection, particularly for signals with low \ac{SNR} (see Supplementary material).

\section{Conclusions}

We analysed the \ac{GSHE} on the waveforms of lensed \acp{GW}. The \ac{GSHE} is a strong field effect that describes the propagation of polarized wave packets. It produces a frequency-dependent delay in the inspiral part of the waveform, while keeping the merger and ringdown relatively unchanged, as shown in~\cref{fig:circ_waveform}. The delay has a characteristic dispersive $1 / f^2$ dependence, mimicking a non-zero graviton mass when $\beta > 0$ and may appear as a violation of Einstein's theory if not properly taken into account.

We identified two promising scenarios for the detection of the \ac{GSHE}. One requires the source and observer to be aligned, leading to highly magnified images with a strong GSHE imprint. In this case, the source could be quite far from the background \ac{BH} ($r_{\rm src}\gg 5 R_S$) and magnification bias facilitates the observation of the signal.
In the second case the source and observer are not aligned and one of the bundles is strongly deflected, grazing the light ring of the background \ac{BH}. These images are usually demagnified and too faint except for sources close to the background \ac{BH}. AGNs and globular clusters, two of the formation channels for stellar-mass binary \acp{BH}, can potentially host such events. 
Although the number of such sources is unknown~\citep{Gerosa2021}, the GSHE provides additional means to investigate their existence. 

In addition to coalescing stellar-mass binaries, the GSHE can also be detected for binaries in the early inspiral phase. This is a prime target for proposed low-frequency detectors in the $\mu$Hz, mHz and dHz bands~\citep{LISA17,LISA19,Gong:2021gvw,Sedda:2019uro,Baibhav:2019rsa,Sesana:2019vho}. 
The dependence of the corrections, Eq.~\eqref{eq:epsilon_SI}, shows that low-frequency \ac{GW} result in comparable signatures for much heavier background \acp{BH}, allowing the \ac{GSHE} to probe central \acp{BH} of galaxies. Migration in these systems could plausibly drive stellar-mass objects towards very small radii \citep{Peng:2021vzr}. The analysis of long-lived inspirals requires extending the framework to moving sources, a subject of future work.
In addition, we expect the \ac{GSHE} signal to be strongest when the \ac{GO} expansion fails, $\epsilon\sim 1$. Exploring this regime requires a framework that combines wave optics and strong gravity \citep{Cardoso:2021vjq,Pijnenburg:2024btj}.

Detecting the \ac{GSHE} can establish an association between stellar-mass binaries and more massive \acp{BH}. 
\ac{GSHE} imprints due to intermediate-mass \acp{BH} provides new means to characterize these elusive objects, complementary to tidal disruption events~\citep{Wen:2021yhz}, fast-moving stars in globular clusters \citep{2024Natur.631..285H} and lensed gamma-ray bursts~\citep{Paynter:2021wmb,Yang:2021wwd,Wang:2021ens}.
If observed in massive \acp{BH}, the \ac{GSHE} will augment the knowledge of AGN binaries derived from their intrinsic parameters, peculiar motion or electromagnetic counterparts, and cross-correlation~\citep{Tagawa:2019osr,Vijaykumar:2023tjg,Morton:2023wxg,Veronesi:2023ugk}.
This information will directly inform the binary formation scenarios and probe their close environment.

Furthermore, a detection of the \ac{GSHE} will also confirm the general-relativistic strong-field effects on the propagation of \acp{GW}, responsible for spin-orbit interactions of the same type as in optics \citep{SOI_review} and condensed matter physics \citep{SHE_review}. Discovering a \ac{GW} source lensed by a \ac{BH} will also provide an exquisite test of alternative gravity theories that produce modifications in regions of high curvature~\citep{Ezquiaga2020,Goyal:2023uvm,Eichhorn:2023iab}. Additionally, investigating the spacetime surrounding the background \ac{BH} might be feasible, for instance, probing superradiant clouds due to ultralight bosons \citep{Brito:2015oca}.
In summary, \ac{GW} observations offer potential for experimental verification of the \ac{GSHE}, providing a test of \acp{GW} propagating in strong gravitational fields and potentially enabling novel applications in astrophysics and fundamental physics.

\section*{Acknowledgements}

The authors thank Lars Andersson, Pedro Cunha, Dan D'Orazio, Francisco Duque, H\'ector Estell\'es, Bence Kocsis, Johan Samsing, Laura Sberna and Jochen Weller for input and discussions, as well as the anonymous referee for constructive comments and recommendations. RS acknowledges financial support from STFC Grant No. ST/X508664/1 and the Deutscher Akademischer Austauschdienst (DAAD) Study Scholarship.

\section*{Data Availability}

The code underlying this article is available at \citep{GSHE_code} and other data will be made available on reasonable request to the authors.

\appendix

\section{Geometrical optics \& Gravitational spin Hall effect complementarity}
\label{supp}

Here, we present some details of the complementarity between the \ac{GO} and \ac{GSHE} observations. To investigate the issue, we generated $\sim 16000$ pairs of trajectories connecting randomly placed sources and observers, assuming $r_{\rm src}=5 R_{\rm s}$ from a \ac{BH} with spin $a=0.99$. 

\begin{figure}
    \centering
    \includegraphics[width=0.48\textwidth]{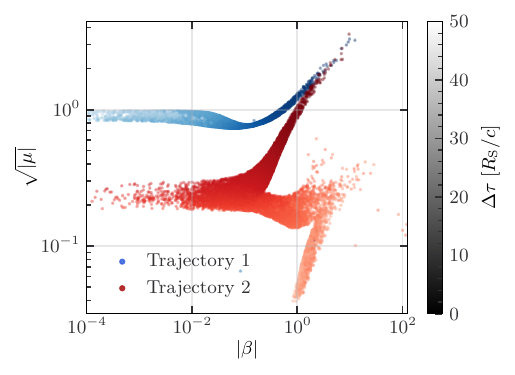}
    \caption{\ac{GO} signal and \ac{GSHE} amplitude for images 1 \& 2 (blue/red), for $r_{\rm src}=5\,R_{\rm s}$, $a=0.99$. Points are color coded by the time delay between \ac{GO} images, with darker values corresponding to closer arrival times.}
    \label{fig:mu_beta}
\end{figure}

Figure \ref{fig:mu_beta} shows the magnification $|\mu_i|$ and \ac{GSHE} amplitude $|\beta_i|$. Positive/negative parity signals (trajectories 1/2) are marked in blue/red and color-coded by the time delay.
A small fraction of the configurations has large $|\mu_i|$ and $|\beta_i|$, with a small time delay between the \ac{GO} trajectories. This corresponds to a close alignment between the source, lens, and observer, similar to Figs.~1 and 2 in the main manuscript. Generic trajectories have a positive parity signal with $\mu_1\sim 1$ (slightly below unity due to gravitational redshift) with a negative parity image with low amplitude $|\mu_{2}|\ll 1$ and a sizeable \ac{GSHE} $\beta_{2}$.

The detectability of multiple \ac{GO} images and \ac{GSHE} signatures depends mainly on the unlensed \ac{SNR} ($\rho_0$). 
We want to distinguish whether a signal is \textit{detectable} (with optimal analysis) and \textit{detected} (using standard techniques). We assume that a \ac{GO} signal is detected if
\begin{equation}\label{eq:GO_detected}
\sqrt{|\mu_i|\left(1-\mathcal{M}(\beta_i)\right)}\rho_0 \geq \rho_{\rm th}\quad \text{(\ac{GO} detected)}\,,
\end{equation}
where $\rho_{\rm th}=8$ is the detection threshold and $\mathcal{M}(\beta_i)$ is the mismatch due to the \ac{GSHE} ~\cite[Sec. II.E]{GSHE_lensing}, which can prevent the identification of a signal. 
In contrast, a signal is detectable for
\begin{equation}\label{eq:GO_detectable}
\sqrt{|\mu_i|}\rho_0 \geq \rho_{\rm th}\quad \text{(\ac{GO} detectable)}\,,
\end{equation}
if the \ac{GSHE} is accounted for. 
We will consider the \ac{GSHE} detected/detectable if $\sqrt{|\mu_i|\mathcal{M}(\beta_i)}\rho_0>1$~\citep[Sec. IV.E]{GSHE_lensing}, in addition to Eqs.~\eqref{eq:GO_detected} and \eqref{eq:GO_detectable}.
For concreteness, we will now assume a lens \ac{BH} with $M=10^4M_\odot$ and a non-spinning equal mass source with total mass $20M_\odot$.

\begin{figure}
    \centering
    \includegraphics[width=\linewidth]{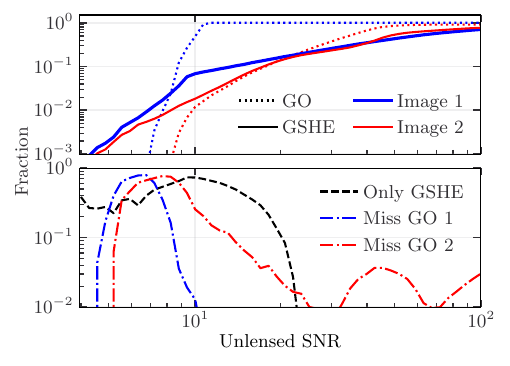}
    \caption{\ac{GO} \& \ac{GSHE} complementarity. 
    \textbf{Top:} fraction of signals with a detection of \ac{GO} (dotted) and potentially detectable GSHE (solid) for $\rho_{\rm th}=8$, binary total mass $20M_\odot$, lens \ac{BH} with $M=10^4M_\odot$ and source at $5 R_{\rm s}$. 
    \textbf{Bottom:}
    Fraction of signals where an \ac{BH} is detectable via GSHE (dashed) and where a \ac{GO} image is missed due to GSHE (dot-dashed).}
    \label{fig:detection_fractions}
    \vspace{10pt}
\end{figure}

Figure \ref{fig:detection_fractions} shows the fraction of signals that satisfy the following conditions as a function of the unlensed \ac{SNR}:
\begin{enumerate}\setlength{\itemsep}{5pt}
\item \ac{GO} signal is detected (dotted)
\item \ac{GO} \& \ac{GSHE} are detectable (solid)
\item \ac{GO} missed due to \ac{GSHE} (dashed dotted).
\item \ac{GSHE} is detectable in signal 1, but signal 2 below threshold (dashed)
\end{enumerate}
For large $\rho_0\gtrsim 50$ both \ac{GO} signals are almost always detected, along with the \ac{GSHE} imprints in most cases. The risk of missing a \ac{GO} image by not accounting for the \ac{GSHE} is low for the negative-parity image ($\lesssim 5\%$) and negligible for the positive-parity image. The high $\rho_0$ situation is expected for next-generation detectors, which are not magnitude limited.

The \ac{GSHE} plays a crucial role for sources near or below the detection threshold $\rho_0\lesssim \rho_{\rm th}$.
The fraction of missed signals can be substantial $\sim 80\%$, as detectable signals represent aligned configurations with high $|\mu_i|$ and $|\beta_i|$ (top right of Fig.~\ref{fig:mu_beta}).
There is also a significant chance ($20-80\%$) that a \ac{BH} near the source can only be detected by measuring $\beta_1\neq 0$ on the positive-parity image, since image 2 is undetectable even when accounting for GSHE (Fig.~\ref{fig:detection_fractions} bottom, dashed line). 
This probability is significant even for moderate \ac{SNR} $\rho_0\lesssim 20$. Incorporating the \ac{GSHE} in the analysis can therefore be important at the current detector sensitivity, where most detected sources have \ac{SNR} close to the detection threshold, and magnification bias may play an important role.%

Our analysis has neglected both birefringent \ac{GSHE} ($\beta_{LR}$) and the fact that multiple families of trajectories can interfere if the time delay is short. Both of these factors increase the mismatch, which reinforces the need for more detailed modeling to test the \ac{GSHE}.
The impact of $\beta_{LR}$ is suppressed by an additional power of $1/f$. However, a fraction of the trajectories have $|\beta_{LR}|\gg |\beta|$.
Ultimately, the impact of birefringent \ac{GSHE} on the waveform depends on the \ac{GW} polarization (i.e. via the inclination angle): we will leave a quantitative analysis for a future study.

\begin{figure}
    \centering
    \includegraphics[width=\linewidth]{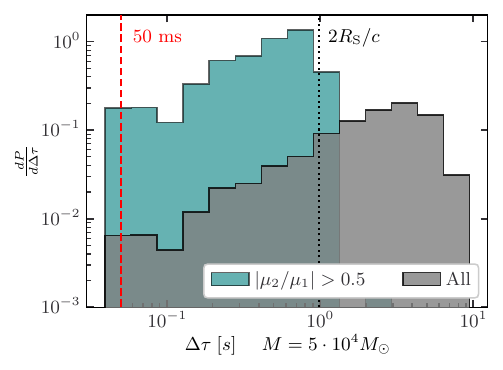}
    \caption{Distribution of the time delays between GO trajectories. Two cases are shown: all trajectories (gray) and those where both GO signals have comparable amplitude (teal): this selects closer source-lens-observer alignments, for which the time delay is smaller. The vertical lines show the time delay in Figs. 1 and 2 of the letter and the characteristic scale.}
    \label{fig:delta_t}
\end{figure}

The probability of overlapping signals depends on the time delay distribution. This is shown in Fig.~\ref{fig:delta_t} for the system considered in the main part of the paper ($M=5\cdot 10^4M_\odot,\, a=0.99, r_{\rm src}=5 R_{\rm s}$), for arbitrary trajectories and those with comparable amplitudes $|\mu_{2}/\mu_{1}|> 0.5$. Overlapping trajectories are generally a small fraction ($\lesssim 0.4\%$ with $\Delta \tau<0.3$s), but become more likely when \ac{GO} signals have a comparable amplitude ($\lesssim 10\%$  with $\Delta \tau<0.3$s and $|\mu_{2}/\mu_{1}| > 0.5$). Accounting for interference will be important to detect and analyze aligned source-\ac{BH}-lens configurations, where both magnifications and \ac{GSHE} are large (e.g. Figs. 1 and 2 in the letter).

We also assumed that the same detection threshold holds for both GO signals. However, once the brighter signal has been identified, a dedicated analysis can find an image with a lower $\rho_{\rm th}$~\citep{Wang:2022ryc} (with or without accounting for GSHE). Other factors, such as source motion~\citep{Zhang:2023cdh} and orbital inclination~\citep{Gondan:2021fpr} cause differences between \ac{GO} signals and need to be taken into account. Our analysis has assumed a source close to an intermediate mass \ac{BH} ($M=10^4M_\odot$, $r_{\rm src}=5\,R_{\rm s}$). Higher/lower \ac{BH} masses will make the \ac{GSHE} less/more relevant (although lighter lenses will require a wave-optics treatment of strong fields, currently not available). Increasing the source's distance lowers the magnification of strongly deflected trajectories, reducing the detectability of generic configurations. However, the \ac{GSHE} remains important for close source-\ac{BH}-observer alignments, which will happen in a fraction of sources $\propto R_{\rm s}/r_{\rm src}$. Given the growing rate of \ac{GW} detections, the \ac{GSHE} opens a new avenue for studying intermediate mass \acp{BH} and their properties.

\bibliographystyle{mnras}
\bibliography{references} 




\bsp	
\label{lastpage}
\end{document}